\documentclass[twocolumn,aps,preprintnumbers,showpacs,floatfix,superscriptaddress]{revtex4}
\usepackage{amsmath,amssymb,bm,epsfig}

\renewcommand\d{\partial}
\newcommand\de{\partial}
\newcommand\x{\bm{x}}
\renewcommand\v{{\bm{v}}}
\newcommand\p{{\bm{p}}}

\newcommand\+{\dagger}

\newcommand{\mudiff}{H}
\newcommand\eone{\epsilon}

\begin{document}

\preprint{INT-PUB 05-18}

\title{Phase diagram of a cold polarized Fermi gas}
\author{D.~T.~Son}
\affiliation{Institute for Nuclear Theory,
University of Washington, Seattle, Washington 98195-1550, USA}
\author{M.~A.~Stephanov}
\affiliation{Department of Physics, University of Illinois, 
Chicago, Illinois 60607-7059,USA}

\begin{abstract}
We propose a phase diagram for a cold polarized atomic Fermi gas with
zero-range interaction.  We identify four main phases in the plane of
density and polarization: the superfluid phase, the normal phase, the
gapless superfluid phase, and the modulated phase.  We argue that
there exists a Lifshitz point at the junction of the normal, the
gapless superfluid, and the modulated phases, and a splitting point
where the superfluid, the gapless superfluid, and the modulated phases
meet.  We show that the physics near the splitting point is universal
and derive an effective field theory describing it.  We also show that
subregions with one and two Fermi surfaces exist within
the normal and the gapless superfluid phases.

\end{abstract}
\date{July 2005}

\maketitle

\section{Introduction}

The Fermi gas in the regime of large scattering
length $a$~\cite{Eagles,Leggett,Nozieres} 
has attracted much interest due to its
universal behavior.  The regime can be achieved in atom traps by using
the technique of Feshbach 
resonance~\cite{Jin,Grimm,Ketterle,Thomas,Salomon}.  Most
attention is focused on systems consisting of two 
components of fermions 
(e.g., two spin components of a spin-$\frac12$ fermion) with equal
number density.  When the effective range $r_0$ is small
compared to the interparticle distance $n^{-1/3}$, where
$n$ is the total
number density, many properties of the system depend
on $n$ and $a$ only through the 
dimensionless diluteness parameter 
\begin{equation}\label{kappa}
  \kappa = -\frac1 {na^3}\,.
\end{equation}
When one varies $\kappa$ the system interpolates between the
Bose-Einstein condensation (BEC) regime and the
Bardeen-Cooper-Schrieffer (BCS) regime.
For all values of $\kappa$ the ground
state is believed to be a superfluid.

In contrast, the case of unequal number density (or unequal chemical
potentials) of the two 
components is much less
understood~\cite{Bedaque,PaoWuYip,Carlson:2005kg}. In the case
of spin-$\frac12$ fermions one refers to a polarized gas.  We
follow this terminology, understanding ``polarized'' in the sense of
asymmetry between the two 
components.

\begin{figure}[t]
\centerline{\epsfig{file=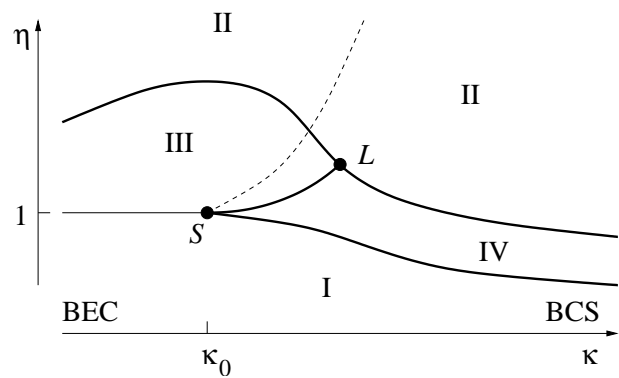,width=0.45\textwidth}}
  \caption[]{The proposed phase diagram in the plane
of the diluteness parameter $\kappa=-1/(na^3)$ and the (scaled) polarization
chemical potential $\eta=H/\Delta_{H=0}$.
I is the unpolarized BEC/BCS phase, II is the normal phase, 
III is the gapless superfluid phase, and IV is
a region of Fulde-Ferrell-Larkin-Ovchinikov phases.  
The dashed line divides phases II and III into
regions with one (on the left) and two (on the right) Fermi surfaces.
Region IV must be
divided into phases with different patterns of breaking of the
rotational symmetry (not shown).}
  \label{fig:vf2}
\vspace{-3ex}
\end{figure}

In this paper we propose a phase diagram for a polarized Fermi gas in
the whole range from the BEC to the BCS regime.  Our proposal is
summarized in Fig.~\ref{fig:vf2}.  
There are two special points on the phase diagram.  Point $S$
(the splitting point) is a point where phases I, III and IV
meet. Point $L$ is a Lifshitz point where II, III and IV meet.
The focus of this paper is on the point $S$. The physics in the
vicinity of this point is of a long-distance, i.e.,
universal, nature and can be reliably 
studied within an effective field theory.

Our proposal for the global structure of the phase diagram
is an educated guess anchored on the following reliable facts:  the 
known phase structure in the BEC and BCS limits; 
the existence of the point $S$, which is at least a local minumum of the
free energy; and the structure of the phase diagram around $S$, studied
in detail in this paper.

We devote the Appendix to the question of the global stability of the
splitting point. We show that quantum Monte Carlo simulations
\cite{Carlson:2005kg} indicate that the superfluid state at
point $S$ is
globally stable. We also argue that mean-field
calculations of the phase diagram will likely miss the point $S$
by significantly overestimating the size of the region occupied 
by phase II.

\section{Qualitative description of the phase diagram}

\subsection{Axes of the phase diagram}

A particular system is
characterized by three parameters: the scattering length $a$ and the
chemical potentials of the two 
components $\mu_\uparrow$ and
$\mu_\downarrow$.  Because of universality (corresponding to rescaling
invariance $a\to e^{-s}a$,
$\mu_i\to e^{2s}\mu_i$) the whole phase diagram can be captured in a
two-dimensional plot.  We introduce the notation
\begin{equation}
  \mu = \frac12(\mu_\uparrow+\mu_\downarrow), 
  \qquad \mudiff = \frac12
(\mu_\uparrow-\mu_\downarrow)\, .
\end{equation}
Then the parameter $\kappa$ on the horizontal axis is defined by 
Eq.~(\ref{kappa})
where $n=n(\mu, a)$ is the density of an unpolarized gas at chemical
potential $\mu$ and scattering length $a$.  Thus $\kappa$ is the
inverse diluteness parameter of an unpolarized system with chemical
potential equal to the average of $\mu_\uparrow$ and $\mu_\downarrow$
and with the same scattering length $a$.

The parameter $\eta$ on the vertical axis is defined as
\begin{equation}\label{eta}
  \eta = \frac{\mudiff}{\Delta(\mu, a)}\, .
\end{equation}
Here $\Delta(\mu,a)$ is the gap in the fermion excitation spectrum of
the unpolarized gas.
We divide $\mudiff$ by $\Delta$ so that $\eta$ is dimensionless.  This is also
convenient because $\eta=1$ is the line dividing phases I and III, as
explained below.

\subsection{Qualitative understanding of the phase diagram}

The general
structure of the phase diagram can be guessed by studying the BEC
and BCS limits first.  In the BEC limit $\kappa\to-\infty$, the
gap in the fermionic excitation spectrum is close to the two-body
binding energy.  The excitations are fermionic quasiparticles carrying charge
$+1$ and $-1$ with respect to the chemical potential $\mudiff$.  When
$\mudiff<\Delta$ or $\eta<1$, the ground state is the same as for
$\eta=0$ since it is energetically unfavorable to create an unbound
fermion on top of the BEC ground state (phase~I).  However, if $\eta>1$, extra
fermions will be created, and the system, for a range of
$\eta$, is a homogeneous mixture of bosonic bound states and
fermions carrying one sign of spin (phase III).  Finally, for
sufficiently 
large~$\eta$ all bosonic bound states disappear and the system is a
completely polarized Fermi gas (phase II).

The ground state in the BCS limit $\kappa\to+\infty$ is also known.  For
$\eta<\eta_1\approx1/\sqrt2$, the ground state remains the BCS
state with zero polarization (I).  For
$\eta_1<\eta<0.754$, the ground state is in one
of the Fulde-Ferrell-Larkin-Ovchinnikov (FFLO)
states~\cite{FuldeFerrell,LarkinOvchinnikov} where Cooper pairs form
with nonzero momentum and the superfluid order parameter varies in space
(IV).  The precise spatial structure of the ground state is
difficult to determine, but it is presumably crystalline.  Finally for
$\eta>0.754$ the system goes to the Fermi liquid phase II.

At small $\eta$, 
the unpolarized BEC phase can continue analytically
into the unpolarized BCS phase.  Indeed, in both regimes the U(1) symmetry
associated with the conservation of the total number of atoms is
spontaneously broken. At large $\eta$, 
the normal phase on the BEC side and that on the BCS side are
essentially the same phase of a
polarized Fermi gas. 
On both sides  the U(1) symmetry is restored.
However, 
at intermediate $\eta$ the phases in the two limits are qualitatively
different: on the BCS side the translational invariance is
spontaneously broken, while on the BEC side it is not,
as already observed in Ref.~\cite{PaoWuYip}.

Under the assumption that the phase diagram contains four
phases which enter the region
of intermediate~$\kappa$ and~$\eta$---I from below, II from above,
III from the left and IV from the right---there are two possibilities.
The first is that there is a phase transition (coexistence) line
separating the boson-fermion mixture phase III and the FFLO phase IV.
The second is that III and IV do not coexist, but instead I and II do,
which is what is found in mean-field theory. In the Appendix we argue that
this is a likely artifact of the mean-field approximation.

Within the first possibility,
the minimal phase diagram therefore should look like 
Fig.~\ref{fig:vf2}.  The line
which separates the mixture phase III from the FFLO phase IV
should start at a
point $S$ located on the boundary of the unpolarized BEC/BCS phase I,
and end at a point $L$ on the boundary of the normal phase II.  The
point $S$ will be called the splitting point, since, as we will show
below, it is the point where the onset transition (from the
unpolarized phase I to the mixture phase III) splits into two first-order
phase transitions.  The point $L$, by the nature of the phases surrounding
it, is a Lifshitz point \cite{ChaikinLubensky}.

In the next section we shall identify the splitting point $S$ and
study the phases surrounding it.  We shall see that the physics around $S$
matches very well with the global picture presented above.

\section{The splitting point}

\subsection{Location of the splitting point}

\begin{figure}[t]
\centerline{  \epsfig{file=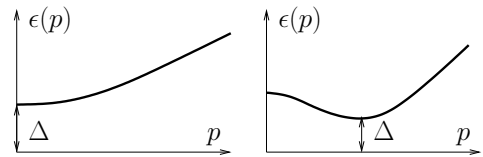,width=.35\textwidth}}
\caption[]{Spectrum of fermionic quasiparticles at $\eta=0$ 
for $\kappa<\kappa_0$ (left) and for $\kappa>\kappa_0$ (right).
The minimum is located at $p=0$ for $\kappa<\kappa_0$ and $p\neq0$ for
$\kappa>\kappa_0$.}
\label{fig:dispersion}
\end{figure}

Begin by
considering the unpolarized gas, i.e., $\eta=0$.  The
dispersion relation of a fermionic quasiparticle
$\epsilon(\p)$ has a finite gap $\Delta$ in the whole range of $\kappa$,
from BEC to BCS. There is, however, a qualitative change in
the location of the minimum of the dispersion curve as illustrated
in Fig.~\ref{fig:dispersion}.
  In the BEC limit
the dispersion curve is $\epsilon(\p) = p^2/(2m)+\Delta$ and
achieves its minimum at $p=0$.  On the other hand, in the BCS limit
the dispersion curve has a minimum at $p\ne0$ (near the Fermi momentum).

One concludes that, as the system crosses over from the BEC to the BCS regime,
\emph{at a certain point $\kappa=\kappa_0$ the minimum of the dispersion
curve shifts from $p=0$ to $p\neq0$}.  In the mean-field theory this 
occurs when the chemical potential vanishes~\cite{Leggett}; we shall only
assume that
this transition occurs smoothly.
Around this $\kappa$ the dispersion
curve has the following behavior near $p=0$:
\begin{equation}\label{epsilonp}
  \epsilon(\bm p) = \epsilon_0 - \alpha p^2 + \beta p^4\,,
\end{equation} 
where $\alpha$ changes sign at $\kappa=\kappa_0$:
$\alpha=\alpha_0(\kappa-\kappa_0)$, $\alpha_0>0$.  
Note that $\Delta=\epsilon_0-\theta(\alpha)\,\alpha^2/(4\beta)$.

At $\eta=0$, the change in the location of the minimum of the
dispersion curve by itself does not signal any phase transition.
However, as we shall see shortly, the splitting point $S$ is located
at $\kappa=\kappa_0$ and $\eta=1$.

The value of $\kappa_0$ cannot be found from effective theory.  One can
determine the sign of $\kappa$ from the shape of the dispersion curve
at infinite scattering length ($\kappa=0$).  According to a recent
quantum Monte-Carlo calculation~\cite{Carlson:2005kg}, at $\kappa=0$
the dispersion is BCS-like, i.e., has a minimum at a nonzero value of
$p$. 
From this
information we conclude that $\kappa_0<0$, i.e., on the left (BEC)
side of the phase diagram in Fig.~\ref{fig:vf2}.

\subsection{Effective field theory near the splitting point}

Let us
take the system at some $\kappa$ close to $\kappa_0$ and increase
$\eta$.  When $\eta$ is close to 1, there are 
fermion quasiparticles whose energy $\epsilon=\epsilon(\bm p)-H$ 
is small.  The momenta of these quasiparticles are also small, 
around the minimum of the dispersion curves.  
At small number density, these quasiparticles are weakly interacting.  
The whole problem is therefore treatable by using a low-energy
effective field theory of quasiparticles (superfluid phonons {\em and\/} the
extra fermions) despite the fact that the original (undressed) particles are 
strongly coupled.

The effective Lagrangian is constrained by two U(1) symmetries.
One of them corresponds to the conservation of the total number of atoms.
The superfluid mode $\varphi$, which we normalize to be half of the phase of
the Cooper pair, transforms under this symmetry as $\varphi\to\varphi+\alpha$. 
The fermion field $\psi$ can always be chosen to be neutral under this symmetry
by multiplying it by an appropriate power of $e^{i\varphi}$.
Another U(1) symmetry corresponds to the conservation of the
difference of the numbers of the two 
components, or polarization. Under this
symmetry $\psi\to\psi e^{i\beta}$, while $\varphi$ is invariant. 
To lowest nontrivial order in
$\varphi$ and $\psi$ and derivatives the Lagrangian obeying these symmetries
is given by
\begin{equation}\label{Leff}
  {\cal L} = \frac{f_t^2}2 \dot\varphi^2
- \frac{f^2}2 (\bm\de\varphi)^2
+ \psi^\dagger [i\de_0 - \eone(-i\bm \de)]\psi
  - g \bm\de\varphi
\frac{i\psi^\dagger  
\raisebox{0.1em}{$\stackrel{\scriptstyle\leftrightarrow}{\bm\d}\,$}
  \psi}{2m} \,.
\end{equation}

The low-energy parameters $f_t$,
$f$ and $g$ are not constrained by the U(1) symmetries.
Their values are not essential for the discussion of the phase
diagram below. However,  for the purpose of
further applications, we point out that they are constrained 
by the Galilean invariance.  
Using the results of 
Refs.~\cite{Greiter:1989qb,Son:2002zn}, one can show that $f_t^2=dn/d\mu$,
$f^2=n/m$, and $g=1$.

Note that the interaction term
$  \psi^\+ \psi \left[ \dot\varphi + (\bm\de\varphi)^2/(2m)\right]$
is allowed, but is negligible when 
the number density of fermion quasiparticles is small,
$\psi^\+\psi\ll n$.  The fermion self-interaction $(\psi^\+
\raisebox{0.1em}{$\stackrel{\scriptstyle\leftrightarrow}\d\,$}\psi)^2$
is also irrelevant.

\subsection{Phases near the splitting point}
\begin{figure}[t]
\centerline{\epsfig{file=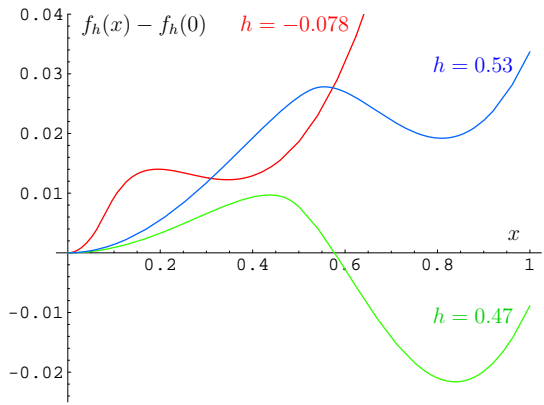,width=0.4\textwidth}}
\caption[]{Function $f_h(x)$ at three representative values of $h$.}
\label{fig:potential}
\end{figure}

When $\alpha>0$ the ground state may
carry nonzero spatial
gradient of $\varphi$ or, in other words, nonzero superfluid velocity
$\v_s =\bm\d\varphi/m$.  
This can be seen, e.g., from the fact that a state with an arbitrarily small
density of fermions and $\v_s=0$ has a negative superfluid
density due to the divergent density of states at the Fermi 
surface~\cite{AronovSpivak}.
Our task is to find, at each value of
$\alpha$ (close to 0), the minimum of the free energy as a function of
$\v_s$.  Using Eqs.~(\ref{epsilonp}) and (\ref{Leff}) we find the fermion 
dispersion relation in the presence of a uniform superfluid flow with velocity
${\bm v}_s$:
\begin{equation}
  \epsilon_v(\p) = \epsilon_0 - \alpha p^2 + \beta p^4 + \v_s\cdot\p 
     -\mudiff .
\end{equation}
All levels with $\epsilon_v(\p)<0$ are filled.  The total
free energy receives contributions from the superfluid flow and from the filled
fermionic energy levels,
\begin{equation}
  F(v_s) = \frac\rho2 v_s^2 +\int\!\frac{d^3\p}{(2\pi)^3}\, 
      \epsilon_v(\p)\, \theta(-\epsilon_v(\p))\,.
\end{equation}
Here $\rho=mn$, and $\theta(x)$ is the step function. Performing the
integration, we find the free energy,
\begin{equation}
  F(v_s) = \frac1{4(15\pi^2)^4}\frac{\alpha^5}{\rho^3\beta^7}
      f_h(x) ,
\end{equation}
where we rescaled $v_s$ and $(H-\Delta)$ by introducing dimensionless
variables
\begin{eqnarray}
  x &=& \sqrt2 (15\pi^2)^2\, \frac{\rho^2\beta^{7/2}}{\alpha^{5/2}}v_s\,,\\
  h &=& 2(15\pi^2)^2\, \frac{\rho^2\beta^4}{\alpha^3}
     (\mudiff-\Delta)\,.
\end{eqnarray}
The shape of $F(v_s)$ at a given $H$ is captured by the function $f_h(x)$,
\begin{equation}\label{f}
  f_h(x) 
  =  x^2 
  -\frac1x \Bigl[ (h+x)^{5/2}\theta(h+x)
  -(h-x)^{5/2}\theta(h-x) \Bigr],
\end{equation}
at the corresponding $h$, which is shown in Fig.~\ref{fig:potential}. 
For $h<h_1\approx-0.067$ the absolute minimum of $f$ is located at
$x=0$.  When $h_1<h<h_2\approx0.502$ the minimum switches to $x\neq0$.  When
$h>h_2$, the minimum switches back to $x=0$.

The change in the location of the minimum of $f$ translates into two
{\em first-order} phase transitions occurring at 
\begin{equation}\label{eta12}
  \eta_{1,2} = 1 + \frac{h_{1,2}}{2(15\pi^2)^2}
  \frac{\alpha^3}{\rho^2\beta^4\Delta}\,.
\end{equation}
The following picture emerges.
At fixed $\kappa>\kappa_0$, as we increase $\eta$ 
from 0 the ground state is an unpolarized 
superfluid (phase I) for $\eta<\eta_1$, 
until a phase transition at $\eta=\eta_1$.
Note that, since $h_1<0$,
the phase transition occurs while $H$ is still below the gap:
$\mudiff_1<\Delta$.

\begin{figure}[ht]
\centerline{  \epsfig{file=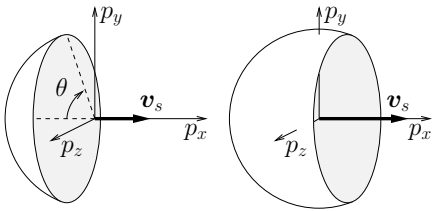,width=.32\textwidth}}
\caption[]{Occupied fermion quasiparticle modes in phase IV near point $S$
occupy a thin spherical cap with curvature radius $p=\sqrt{\alpha/(2\beta)}$.
The angle $\theta$
increases from $80.4^\circ$ at $\eta=\eta_1+0$ to $127.3^\circ$
at $\eta=\eta_2-0$.}
\label{fig:cap3}
\end{figure}

\begin{figure}[ht]
\centerline{  \epsfig{file=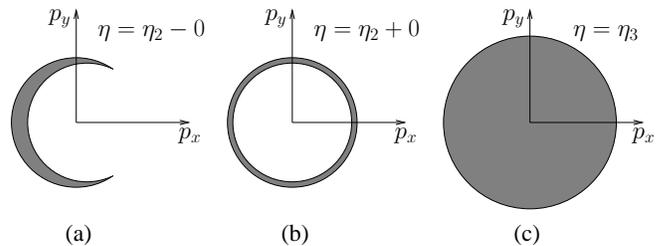,width=.48\textwidth}}
\caption[]{Occupied quasifermion modes ($p_z=0$ cross section) near point $S$
  in
phase IV (a) and in phase III (b),(c).}
\label{fig:shell-sphere}
\end{figure}

For
$\eta_1<\eta<\eta_2$ the ground state has nonzero superfluid
velocity $v_s$ (phase IV). It also contains fermionic quasiparticles 
filling an asymmetric
region in the phase space 
[Figs.~\ref{fig:cap3} and \ref{fig:shell-sphere}(a)].
The currents carried by the superfluid flow and
by fermion quasiparticles cancel each other so that all
currents (both total number and polarization) in the ground state are
zero.

When $\eta>\eta_2$ the superfluid velocity switches back to zero.
However, now $\mudiff>\Delta$, 
so there is a finite density of
fermionic quasiparticles
(phase III).  At $\eta=\eta_2+0$
the fermions fill a thin spherical shell
 centered at $\p=0$ [Fig.~\ref{fig:shell-sphere}(b)].  As $\eta$
increases, the shell thickens and at
\begin{equation}
  \eta = \eta_3 \equiv \frac{\epsilon_0}\Delta = 
  1+ \frac{\alpha^2}{4\beta\Delta}\,,
\end{equation}
or $\mudiff=\epsilon_0$, the shell turns into a solid ball 
[Fig.~\ref{fig:shell-sphere}(c)]. This occurs at the
dashed line on Fig.~\ref{fig:vf2}.

\section{Global topology of the phase diagram and concluding remarks}


In the region $\eta_1<\eta<\eta_2$ where $v_s\neq0$, the condensate
function varies as $e^{iqx}$ with $q\equiv 2mv_s\to0$.  This
has the same structure as
the Fulde-Ferrell phase~\cite{FuldeFerrell} in the
BCS regime.  
Therefore, it is natural to expect that the region 
$\eta_1<\eta<\eta_2$ near $S$ is continuously connected to the 
FFLO region on the BCS side, as in Fig.~\ref{fig:vf2}.

In the BCS limit, the ground state is expected to be a crystal where
$\Delta(\x)$ is a superposition of several plane
waves~\cite{Bowers:2002xr}.  However,
$\Delta(\x)$ has to be a single plane wave near $S$.  Indeed, $v_s$ is small
($v_s\to0$ as one approaces $S$), while the healing length is
finite at $S$.  If one tried to superimpose two plane waves to make a
standing wave $\cos(qx)$, the absolute value of the condensate would
vary on length scales much larger than the healing length,
costing a large amount of energy. Therefore, one concludes that the FFLO region
of the phase diagram is further subdivided into smaller regions with
different patterns of breaking of rotational and translational
symmetries.

The line separating the gapless superfluid phase III and the phase
with nonzero $v_s$ IV should end in a Lifshitz point $L$.  
It is known that in
the vicinity of a Lifshitz point, the transition between the phase
with spatially varying order parameter and that with constant order
parameter {\em can\/} be second order when the broken symmetry is
continuous~\cite{ChaikinLubensky}.  If this is the case, then
the first-order line $\eta=\eta_2(\kappa)$ should become second
order at some value of $\kappa$ before the point $L$.  Another
alternative is that the entire line from $S$ to $L$ is first order.

In the region between $\eta=\eta_2(\kappa)$ and
$\eta=\eta_3(\kappa)$, but below the transition to the normal
state (part of region III below the dashed curve on
Fig.~\ref{fig:vf2}), 
the spectrum of low-energy excitations consists of the
superfluid boson and the fermionic quasiparticles with momenta around
\emph{two} Fermi surfaces [the inner and the outer surfaces of the spherical
shell in Fig.~\ref{fig:shell-sphere}(b)].  
This is the Sarma phase~\cite{Sarma},
which is unstable in the BCS limit but, as we found, is stable near the
splitting point.  This is interesting since previous examples of
stabilized Sarma, or breached pairing, phase require quite special
conditions~\cite{Forbes:2004cr}.

Figure~\ref{fig:vf2} describes the system at zero
temperature $T$. At small $T\ne0$ the phase diagram undergoes 
obvious modifications.
Since the phases I and III have the same symmetry, the transition between
them becomes an analytic crossover at $T\ne0$. 
Similarly, the
dashed line on Fig.~\ref{fig:vf2} should turn into a crossover at
$T\ne0$. On the contrary, the phases II and III must remain separated by a
phase transition because the U(1) symmetry is
spontaneously broken in III and restored in II.  Likewise, since the
rotational symmetry undergoes spontaneous breakdown in phase IV, this
phase must remain separated from all other phases by a phase
transition.



The results obtained in this paper should be interesting for other
physical systems with asymmetric fermion pairing, for example for 
the cores of neutron
stars~\cite{Alford:2003fq} or QCD at finite baryon and isospin
chemical potentials \cite{Son:2000xc}. We defer this investigation to
future work.


\acknowledgments

The authors thank A.~Bulgac, A.~Kryjevski, H.-W.~Hammer, and B.~Spivak for
discussions.  This work was supported by DOE grants No.\ DE-FG02-00ER41132
and No.\ DE-FG0201ER41195 and the Alfred P.~Sloan Foundation.

\appendix*

\section{Global stability of the splitting point}

Near the splitting point we 
determined the minima of the free energy with respect to small
perturbations around the superfluid state. By its nature,
the effective field theory we applied cannot assert
that the minima we study are global, i.e., stable with respect
to large deviations from the state we considered.
We must therefore employ different methods to assess the
global stability of the point $S$.
It is an important question since, as we show below, a
mean-field approximation will likely show the normal, not superfluid, phase
to be the globally stable state at the point $S$. Since the mean-field
approximation is not a controllable approximation 
in the relevant regime
$|\kappa|\lesssim 1$, other approaches are necessary 
to determine the global structure of the phase diagram.

Since the point $S$ lies on the line
$\eta=H/\Delta=1$,
the relevant comparision is between the pressure of the
superfluid state  $P_{\rm SF}(\mu,H)$ at $H=\Delta$ 
and the pressure of the normal state $P_{\rm N}(\mu,H)$ at the
same $H$.  We know that the normal state wins in the BCS limit, and the
superfluid state wins in the BEC limit. Thus, there is  a
crossover at some value of $\kappa_c$ 
\begin{equation}\label{kappac}
  P_{\rm SF}(\mu,\Delta)=P_{\rm N}(\mu,\Delta)\qquad\mbox{at}\ \kappa=\kappa_c.
\end{equation}
Assuming  there
is only one such crossover point, the stability of the point $S$ requires that 
$\kappa_0<\kappa_c$, i.e., that the splitting point $S$ occurs 
while the superfluid state is still globally stable, as in Fig.~\ref{fig:vf2}.

Although no direct determination of $\kappa_c$ has been made to date,
Carlson and Reddy~\cite{Carlson:2005kg} have performed the pressure
comparision at the unitarity point $\kappa=0$, using a quantum Monte
Carlo method. We shall now show that these results imply
$\kappa_c\approx0$, as well as that $\kappa_0<0$, 
and thus that the splitting point
is globally stable.

At the point $\kappa=0$ the only dimensionful
parameters are the chemical potentials $\mu$ and $H$.  
The pressure of the superfluid
phase is the same for all $H\in(0,\Delta)$ due to the gap, 
and, by dimensionality, is 
proportional to the pressure of a free two-component unpolarized 
Fermi gas at the same chemical potential,
\begin{equation}
  P_{\rm SF}(\mu, \Delta) = P_{\rm SF}(\mu, 0)
= \frac{P_{\rm free}(\mu)}{\xi^{3/2}}
\end{equation}
where $\xi$ is a universal dimensionless constant, conventionally 
defined as the
ratio of the energy of the gas at unitarity and the energy of the
noninteracting gas at the same density.
$P_{\rm free}(\mu)= 2(2m)^{3/2}/(15\pi^2)\,\mu^{5/2}$.

To find the  pressure of the normal phase, we first notice that
$\mu_\downarrow=\mu-\Delta<0$ since $\Delta>\mu$ at unitarity (both in
mean-field and in quantum Monte Carlo methods).  Therefore the
normal phase is a completely polarized (single-component---hence
a factor 1/2 below) Fermi gas with
$\mu_\uparrow=\mu+\Delta$, whose pressure is
\begin{equation}
  P_{\rm N}(\mu,\Delta) = \frac12 P_{\rm free}(\mu_\uparrow)
=\frac12 \left(1+\frac\Delta\mu\right)^{5/2}
  P_{\rm free}(\mu)
\end{equation}

The ratio of the pressures of the superfluid and the normal phases is
therefore expressed in terms of two dimensionless quantities, $\xi$
and $\Delta/\mu$,
\begin{equation}\label{Pratio}
  \frac{P_{\rm SF}(\mu,\Delta)}{P_{\rm N}(\mu,\Delta)} = 
  \frac2{\xi^{3/2}} \left(1+\frac\Delta\mu\right)^{-5/2}
\quad\mbox{at}\ \kappa=0.
\end{equation}
The Monte Carlo simulation of Ref.~\cite{Carlson:2005kg}, finds 
$\xi\approx0.42$ and $\Delta/\mu\approx1.2$. Substituting into
Eq.~(\ref{Pratio}) one finds the ratio to be consistent with
1 within errors~\cite{Carlson:2005kg}. 
Comparing to Eq.~(\ref{kappac}) we conclude that $\kappa_c\approx0$.

Carlson and Reddy~\cite{Carlson:2005kg} also found that
at $\kappa=0$ the fermion dispersion curve has a miminum at nonzero
momentum, which means that the point $\kappa=0$ is on the
BCS side of the splitting point, i.e., $\kappa_0<0$.
Since $\kappa_0<0\approx\kappa_c$, we can
conclude that the superfluid phase is still globally stable 
at the splitting point.

On the other hand, if we substituted into
Eq.~(\ref{Pratio}) the mean-field values $\xi\approx0.59$,
$\Delta/\mu\approx1.16$ \cite{PapenbrockBertsch}, 
we would find the ratio to be 0.64, i.e., the
superfluid phase would appear to lose by a large margin at $\kappa=0$.
Therefore it is likely that in a mean-field calculation of
the phase diagram the splitting point will be
overshadowed by the normal phase. Instead of Fig.~\ref{fig:vf2},
in the regime $|\kappa|\alt1$ one would then observe a single first-order 
transition (phase co-existence) line separating phases I and II.
Qualitatively, it should be expected that a variational calculation
overestimates the energy of the superfluid phase (i.e., the value of $\xi$).
It is an interesting problem to see if the mean-field theory 
\cite{Bedaque,PaoWuYip} can be improved
in the regime $|\kappa|\alt1$ to describe the phase diagram
correctly.


A calculation of the phase diagram  using a two-channel
Hamiltonian was reported recently~\cite{SheehyRadzihovsky}. 
It relies on a mean-field approximation which becomes controllable
in the weak channel coupling (narrow resonance) limit.
It should be borne in mind that in this
limit the effective range of the interaction
diverges: $r_0\gg n^{-1/3}$.  This limit~\cite{Schwenk:2005ka}
is therefore opposite to the
zero-range limit we consider in this paper.

\end{document}